\documentstyle[prl,aps]{revtex}

\begin{document}
\twocolumn[\hsize\textwidth\columnwidth\hsize\csname %
@twocolumnfalse\endcsname

\title{\bf Exact Results for Kinetics of Catalytic Reactions}

\author{L. Frachebourg and P. L. Krapivsky}
\address{Center for Polymer Studies and Department of Physics}
\address{Boston University, Boston, MA 02215, USA}

\date{\today}
\maketitle
\begin{abstract}

  The kinetics of an irreversible catalytic reaction on a
  substrate of arbitrary dimension is examined.
  In the limit of infinitesimal reaction rate (reaction-controlled limit),
  we solve the dimer-dimer surface
  reaction model (or voter model) exactly in arbitrary dimension $D$.
  The density of reactive
  interfaces is found to exhibit a power law decay for
  $D<2$ and a slow logarithmic decay in two dimensions.
  We discuss the relevance of these results for the monomer-monomer
  surface reaction model.

\smallskip
{PACS numbers:  05.40.+j, 68.10.Jy, 82.20.Mj}
\end{abstract}
\vspace{3mm}
]

In modeling heterogeneous catalysis \cite{cata}, the monomer-monomer
surface reaction model plays an important role, at least from the
theoretical point of view since an appealing simplicity of this model
allows one to examine several issues analytically.  In particular,
investigations of the monomer-monomer model clarified the role of
fluctuations \cite{wicke,ziff,ziff1,meakin,ben,exact1,exact2},
interfacial roughening \cite{kang},  diffusion
of the adsorbants \cite{diff}, and surface disorder \cite{laur}.
In the simplest situation (no diffusion, no disorder, {\it etc.}),
it was found that single-species clusters grow with time when the
dimensionality $D$ of the substrate is sufficiently small,
$D\leq 2$. However, the details of the coarsening like the decay rate
of the density of reactive interfaces remain uncertain in two dimensions,
-- simulations \cite{meakin,evans1,evans2} revealed a very slow
decay which could be logarithmic or power law with a
small exponent.  In this paper, we clarify these questions by computing
analytically kinetic
characteristics of an idealized version of the monomer-monomer model,
the {\it voter model}. We then expand these results and
perform numerical simulations for the full model.

The monomer-monomer surface reaction process can be schematically
represented by the following kinetic steps:
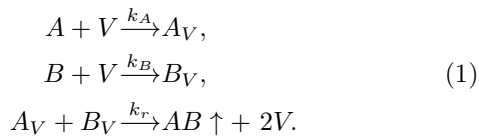
\begin{eqnarray}
A+V & {\buildrel k_A
\over \longrightarrow} & A_{V}, \nonumber \\
B+V & {\buildrel k_B
\over\longrightarrow} & B_{V}, \\
A_{V}+B_{V} &{\buildrel k_r\over \longrightarrow}
& AB\uparrow +\,\, 2V. \nonumber
\label{scheme}
\end{eqnarray}
$A$ and $B$ particles impinge upon a surface, with
respective rates $k_A$ and $k_B$, and adsorb onto vacant sites $V$ to
form a monolayer of adsorbed particles, $A_V$ and $B_V$.
Nearest-neighbor pairs of dissimilar adsorbed particles, $A_VB_V$,
react and desorb with rate $k_r$, leaving behind two vacancies.
For $k_A\ne k_B$, the adsorption imbalance
leads to the quick saturation of the surface by the majority species.
For $k_A=k_B$ and for dimensions $D\leq 2$, there is a fluctuation-induced
coarsening of the surface into growing $A$ and $B$ adsorbed islands.
This nontrivial case of equal adsorption rates will be considered
in the following.
Furthermore, in theoretical analysis we will restrict ourselves to the
{\it reaction-controlled} limit, $k_r\ll k_A=k_B$, which was found to
provide qualitatively the same behavior as the general case \cite{ziff1,ben}
but more amenable to theoretical treatment.

In the reaction-controlled limit, the substrate quickly becomes
completely covered and then stays covered forever, since in units
of the typical, {\it i.e.} adsorption, time interval unoccupied sites
are refilled instantaneously.  The kinetics of the monomer-monomer
surface reaction model is conveniently described
by a mapping onto the Ising model with mixed zero-temperature
voter dynamics and infinite-temperature Kawasaki dynamics \cite{exact2}.
Remember that in the voter model \cite{lig}, sites have
two opinions which can be marked by $A$ and $B$.  Each site keeps its
opinion some time interval, distributed exponentially with
characteristic time $\tau$, and then assumes an opinion of
a randomly chosen neighboring site.  If a site is surrounding by similar
sites it does not change its opinion and therefore the voter dynamics
is zero-temperature in nature.  In the original monomer-monomer
model, the reacting neighboring sites, $A_V$ and $B_V$,
desorb and then unoccupied sites immediately refilled by $A_VB_V$
(no reaction), $A_VA_V$ or $B_VB_V$ (voter dynamics),
or $B_VA_V$ (Kawasaki exchange dynamics whose
effective temperature is infinite since the reaction rate does not depend on
the content of neighboring sites).  Thus the voter model can be
considered as an idealized variant of the monomer-monomer model.

Interestingly, the voter model can be {\it exactly} mapped onto the
reaction-controlled limit of the dimer-dimer surface reaction
model \cite{evans2}.  Indeed, in the dimer-dimer model an empty pair
that appears after the reaction event,
$A_{V}+B_{V} \to AB\uparrow +2V$,  is refilled by an $A_2$ or a $B_2$
dimer, so the resulting dynamics is identical to the voter model
dynamics.  However, natural initial states for the voter model and
the dimer-dimer surface reaction model are different. For the voter
model, a lattice completely covered by $A$ and $B$ monomers without
correlations in initial positions provides a reasonable initial
condition.  For the dimer-dimer model, if we start from an empty lattice
and fill it by random sequential adsorption of dimers one cannot
reach a fully covered lattice and instead approach a so-called
jammed state with some single-cite vacancies\cite{rsa}; this jammed
state provides natural initial condition for the dimer-dimer model.
In the following, we will consider only completely covered initial states.

A remarkable feature of the voter model is its solvability.  It is
evident in one dimension where the voter model is identical to the
kinetic Ising model with zero-temperature Glauber dynamics \cite{lig,glaub}.
Surprisingly, the voter model can be solved in arbitrary dimension and
thus appears to be one of a very few models which are solvable in any
$D$.   Moreover, the voter model can be solved on arbitrary lattice
although for simplicity we will focus on a (hyper)cubic lattice.
To reveal the solvability of the voter model,
the spin formulation is convenient. Identifying $A$'s
($B$'s) with $+$ ($-$) spins so that the state of the substrate is
described by $S\equiv[S_{{\bf k}}], ~{\bf k}=(k_1,\ldots, k_D)$,
one can verify that the spin-flip rate,
$W_{{\bf k}}(S)\equiv W(S_{{\bf k}}\to -S_{{\bf k}})$, reads
\begin{equation}
W_{{\bf k}}(S)={1\over \tau}
\left(1-{1\over 2D}S_{{\bf k}}\sum_{{\bf e}_i}S_{{\bf k+e}_i}
\right).
\label{rate}
\end{equation}
Here the sum in the right-hand side runs over all $2D$ nearest neighbors
and $\tau$ defines the time scale of the process. In the following,
we will set $\tau=4/D$ to simplify numerical factors in equations
for correlators.  The probability distribution
$P(S,t)$ satisfies the master equation
\begin{equation}
{d\over dt}P(S,t)=\sum_{{\bf k}}\left[W_{{\bf k}}(S^{{\bf k}})
P(S^{{\bf k}},t)-W_{{\bf k}}(S)P(S,t)\right],
\label{master}
\end{equation}
where the state $S^{{\bf k}}$ differs from $S$ only at the site {\bf k}.
One can then derive a set of differential equations for the spin
correlation functions
\begin{equation}
\langle S_{{\bf k}}\ldots S_{{\bf l}}\rangle
\equiv \sum_{S}S_{{\bf k}}\ldots S_{{\bf l}}P(S,t).
\label{corr}
\end{equation}
For the single-body correlation functions we get \cite{exact2}
\begin{equation}
4{d\over dt}\langle S_{{\bf k}}\rangle=
\Delta_{{\bf k}}\langle S_{{\bf k}}\rangle.
\label{1body}
\end{equation}
Here $\Delta_{\bf k}$ denotes a difference Laplace operator,
\begin{equation}
\Delta_{\bf k}\langle S_{\bf k}\rangle=-2D\langle S_{\bf k}\rangle
+\sum_{{\bf e}_i}\langle S_{{\bf k+e}_i}
\rangle.
\label{lap}
\end{equation}
For the two-body correlation functions one has \cite{exact2}
\begin{equation}
4{d\over dt}\langle S_{\bf k}S_{\bf l}\rangle=
\left(\Delta_{\bf k}+\Delta_{\bf l}\right)
\langle S_{\bf k}S_{\bf l}\rangle.
\label{2body}
\end{equation}
Similar equations can be written for higher-body correlation functions.
An important feature of these equations which allows an analytical
treatment is their recursive nature, -- to solve for $n$-body correlation
function one does not need the higher correlation functions.  The
structure of equations for correlators is similar to the one that arises
in the one-dimensional kinetic Ising model with zero-temperature Glauber
dynamics \cite{glaub} which is equivalent to the voter model in 1$D$.

The general solution to Eqs.~(\ref{1body}) reads
\begin{equation}
\langle S_{{\bf k}}\rangle=e^{-Dt/2}\sum_{\bf l}\sigma_{\bf l}
I_{{\bf k}-{\bf l}}(t/2).
\label{1sol}
\end{equation}
Here $I_{\bf k}(x)$ is the shorthand notation for the multi-index
Bessel function, $I_{\bf k}(x)=\prod_{1\leq j\leq D}I_{k_j}(x)$,
with $I_n$ being the usual modified Bessel function,
and $\sigma_{\bf l}=\langle S_{\bf l}\rangle(t=0)$.

Although the evolution rules of the voter model do not preserve locally
the densities of $A$'s and $B$'s,
Eq.~(\ref{1sol}) shows that
$\sum_{\bf k}\langle S_{{\bf k}}\rangle=\sum_{\bf l}
\sigma_{\bf l}$, the total densities are conserved.
Note, however, that for any {\it finite} substrate intuitively more appealing
behavior emerges: The effect of fluctuations leads to saturation, {\it i. e.}
all voters eventually share the same opinion thereby stopping the dynamics.

To find the two-body correlation functions, we first make a
simplifying assumption that the initial state is translationally
invariant. Then $\langle S_{\bf k}S_{\bf l}\rangle$ will depend only on
${\bf m}={\bf k-l}$ for $t=0$. Clearly, this holds
for later times, too.  In this situation, the shorthand notation
$R_{\bf m}=\langle S_{\bf k}S_{\bf l}\rangle$ will be used.
Thus for the translationally invariant initial conditions,
Eqs.~(\ref{2body}) simplifies to the lattice diffusion equation
\begin{equation}
2{d\over dt}R_{{\bf m}}=\Delta_{\bf m}R_{\bf m},
\label{2reduced}
\end{equation}
which should be solved subject to the boundary condition
\begin{equation}
R_{{\bf 0}}(t)=1,
\label{source}
\end{equation}
since $R_{{\bf 0}}=\langle S^2_{{\bf k}}\rangle\equiv 1$.
It is natural to choose an uncorrelated initial state,
\begin{equation}
R_{{\bf m}}(t=0)=0, \quad {\rm for}\quad {\rm all}\quad
{\bf m}\ne {\bf 0}.
\label{uncor}
\end{equation}
Eq.~(\ref{2reduced}) is identical to  Eq.~(\ref{1body}) up to a numerical
factor.  Therefore, if we forget for a moment about the boundary
condition of Eq.~(\ref{source}),  we can use Eq.~(\ref{1sol})
to get $\tilde R_{{\bf m}}=e^{-Dt}I_{\bf m}(t)$.  However,
$\tilde R_{\bf 0}=\left[e^{-t}I_{0}(t)\right]^D$ disagrees with
the boundary condition.
To remove this discrepancy it is useful to consider the initial-value
problem, Eqs.~(\ref{2reduced})--(\ref{uncor}), as the problem with a
localized constant source at the origin and therefore to look for the
solution of the form
\begin{equation}
R_{\bf m}=e^{-Dt}I_{\bf m}(t)+
\int_0^t d\tau J_D(t-\tau)e^{-D\tau}I_{\bf m}(\tau).
\label{look}
\end{equation}
Mathematically, Eq.~(\ref{look}) is a linear combination of exact
solutions to Eq.~(\ref{2reduced}) and therefore (\ref{look}) also solves
the linear Eq.~(\ref{2reduced}).  Physically,  Eq.~(\ref{2reduced})
corresponds
to the case of initial source $R_{\bf 0}|_{t=0}=1$ at the origin,
supplemented by an additional input $J_D(\tau)d\tau$ which is added into
the origin during the time interval $(\tau, \tau+d\tau)$ to keep
the overall density at the origin unchanged, $R_{\bf 0}(t)\equiv 1$.
Thus the input strength $J_D$ obeys
\begin{equation}
1-\left[e^{-t}I_0(t)\right]^D=
\int_0^t d\tau J_D(t-\tau)\left[e^{-\tau}I_0(\tau)\right]^D.
\label{J}
\end{equation}
The Laplace transform of the strength,
$\hat J_D(\lambda)=\int_0^\infty dt e^{-\lambda t}J_D(t)$,  can be expressed
through the Laplace transform $\hat T_D(\lambda)$ of the function
$T_D(t)=\left[e^{-t}I_0(t)\right]^D$,
\begin{equation}
\hat J_D(\lambda)=-1+{1\over \lambda\hat T_D(\lambda)}.
\label{hJ}
\end{equation}
Making use of the integral representation \cite{bender},
\begin{equation}
I_0(t)={1\over 2\pi}\int_0^{2\pi}dq~e^{t\cos q},
\label{int}
\end{equation}
we express $\hat T_D(\lambda)$ through the so-called Watson integrals,
\begin{equation}
\hat T_D(\lambda)=\int_0^{2\pi}{d^D{\bf q}\over (2\pi)^D}
{1\over \lambda+D-\sum_{1\leq j\leq D}\cos q_j}.
\label{watson}
\end{equation}
Combining these finding yields
\begin{equation}
  \label{j1}
\hat J_1(\lambda)=\sqrt{{\lambda+2\over\lambda}}
\end{equation}
in $1D$ and
\begin{equation}
  \label{j2}
\hat J_2(\lambda)=-1+\pi{\lambda+2\over 2\lambda}
K^{-1}\left({2\over\lambda+2}\right)
\end{equation}
in $2D$. In Eq.~(\ref{j2}), $K(x)$ is the complete
elliptic integral of the first kind,
$K(x)=\int_0^{\pi/2}d\theta(1-x^2\sin^2\theta)^{-1/2}$.
The final expressions for $J_D(t)$ are given by the inverse
Laplace transform.

{}From the two-body correlation functions,
physically interesting quantities can be found.
One such quantity, the concentration $C_{AB}(t)$ of reactive interfaces,
or nearest-neighbor adsorbed $AB$ pairs,  is given by
$C_{AB}=(1-R_{{\bf e}_i})/2$.
A straightforward computation gives
\begin{eqnarray}
& & C_{AB}(t) = {1\over 2}e^{-Dt}I_0^{D-1}(t)[I_0(t)-I_1(t)]\nonumber \\
& & \qquad + {1\over 2}
\int_0^t d\tau J_D(t-\tau)e^{-D\tau}I_0^{D-1}(\tau)
[I_0(\tau)-I_1(\tau)].
\label{AB}
\end{eqnarray}

The long-time behavior is obtained by analyzing the low-$\lambda$
limit. From Eqs.~(\ref{watson}) and (\ref{hJ}) we get
\begin{equation}
\hat J_D(\lambda)\sim\cases
                      {\lambda^{-D/2} & $D<2$ \cr
                       \lambda^{-1}\ln^{-1}(1/\lambda) & $D=2$ \cr
                       \lambda^{-1} & $D>2$ \cr} \qquad \lambda \to 0
\label{l-asymp}
\end{equation}
which imply
\begin{equation}
J_D(t)\sim\cases
{t^{-1+D/2} & $D<2$ \cr
(\ln t)^{-1} &  $D=2$ \cr
1 & $D>2$ \cr}\qquad \qquad t \to \infty
\label{t-asymp}
\end{equation}

Combining Eqs.~(\ref{AB}) and(\ref{t-asymp}), and the asymptotic
relations $I_0(t)\simeq I_1(t) \simeq e^t/\sqrt{2\pi t},
{}~I_0(t)-I_1(t)\simeq e^t/\sqrt{8\pi t^3}$ \cite{bender} one finds
the asymptotic behavior for the concentration of reactive interfaces
\begin{equation}
C_{AB}(t)\sim\cases
{t^{-1+D/2} & $D<2$ \cr
(\ln t)^{-1} &  $D=2$ \cr
a-bt^{-D/2} & $D>2$ \cr}\qquad t \to \infty
\label{asymp}
\end{equation}
Thus in the long-time limit $C_{AB}\to 0$ when $D\leq 2$,
{\it i.e.} the coarsening, takes place for low dimensional substrates
while for $D>2$ single-species domains do not arise.

In the borderline two dimensional case the coarsening occurs
but the concentration of reactive interfaces decreases very slowly.
In the recent work \cite{evans2}, the decay of  $C_{AB}(t)$
in the voter model has been studied numerically.
Fitting data by power-law and logarithmic forms,
$C_{AB}\sim t^{-\omega}$ and $C_{AB}\sim [\ln t]^{-\sigma}$,
respectively, the effective exponents $\omega\approx 0.096$ and
$\sigma\approx 0.59$ have been observed.
The theoretical asymptotic value $C_AB(t)$ computed from Eq.~(\ref{AB}) is
\begin{equation}
  \label{exact}
C_{AB}(t) = {\pi\over 2\ln(t)+\ln(256)}+{\cal O}\left({\ln t\over t}\right)
\end{equation}
and therefore the asymptotic state arises on a very late stage
which has not been reached in simulations \cite{meakin,evans1,evans2}.
In \cite{evans2}, Evans and Ray did simulations
for times $t\leq 1500 \tau$.  However, trying to fit the exact result
(\ref{exact}) with $C_{AB}\sim [\ln t]^{-\sigma}$,  one can get
at the very best the value $\sigma=0.72$ when $t\approx 1500\tau$.

Turn now to the monomer-monomer surface reaction model in
the reaction-controlled limit. The voter model solution of
Eq.~(\ref{1sol}) is still valid for the monomer-monomer model since
addition of the infinite-temperature Kawasaki dynamics to the voter
model dynamics just results in a change of time scale $\tau$ to
$\tau/2$ in Eq.~(\ref{1body})\cite{exact2}.  Eq.~(\ref{2body})
for $|{\bf{k}-\bf{l}}|>1$ also maintains its form, up to
the replacement $\tau$ by $\tau/2$, while for $|{\bf{k}-\bf{l}}|=1$,
Eq.~(\ref{2body}) undergoes more significant change\cite{exact2}.
However, in the long-time limit the growth of the characteristic space
scale makes the underline lattice structure less and less important
and hence the difference with the voter model behavior should decrease
with time.   So one can expect that the main difference between the
concentration $C_{AB}(t)$ of the voter model and the monomer-monomer
model lies in the time scale difference.  This suggests that
the asymptotic state for the monomer-monomer model is established
later than for the voter model giving rise to a smaller effective
exponent $\sigma$. This hypothesis is comforted by our simulations
where, for longer times than in previous simulations
\cite{meakin,evans1,evans2}, we have always found an effective
exponent $\sigma$ closer to unity.

We have performed numerical simulations for the monomer-monomer model
on a square lattice of size $10^3\times 10^3$ in time interval
$t\leq 10^5 \tau$.  Both the size of the system and time span of the
simulations were significantly bigger than in previous
studies\cite{meakin,evans1,evans2}.
We have found $\sigma\approx 0.62$  which is to be compared with
$\sigma\approx 0.51$ for times $t\leq 1500 \tau$\cite{evans2}.
We believe that in a truly asymptotic regime the value $\sigma=1$
will appear.  Moreover, we think that our results for the kinetics of
the monomer-monomer model of catalysis the reaction-controlled limit
are qualitatively valid for arbitrary reaction rates.
In particular, for the opposite extreme, {\it i.e.} for the
{\it adsorption-controlled} limit ($k_A=k_B\ll k_r$), we have
observed $\sigma\approx 0.42$ instead of $\sigma\approx 0.33$
\cite{meakin} and $\sigma\approx 0.26$ \cite{evans2}. Large time
simulations are again closer to the true asymptotic regime where
we think, $(\ln t)^{-1}$ behavior should be recovered.

In summary, for the voter model in arbitrary dimension we have found
the exact expression for the two-body correlation functions.  In the
most interesting two dimensional case our exact solution reveals,
on the language of the catalysis model, that the density of reactive
interfaces exhibits inverse logarithmic decay. It would be very
interesting to find exact solutions for higher-body correlators.
Given the fact that one dimensional voter model is {\it completely}
solvable\cite{fel}, exact results for low-body correlators in arbitrary
dimension can indicate on the complete solvability of the voter
model for any $D$.  Another direction of further research is
to consider the dimer-dimer model of catalysis in the
adsorption-controlled limit.  Contrary to the monomer-monomer model,
where there is some evidence that the qualitative behavior of the
system is the same in the reaction- and adsorption-controlled limit,
the behavior of the dimer-dimer model should be very different in
these two limiting cases. In the adsorption-controlled limit,
an infinite number of adsorbing states can play a significant role
in the dynamics (existence of adsorbing states is evident,
{\it e.g.} in one dimension any state with $A$- and  $B$-islands
separated by single empty sites will be an adsorbing state).
Note that there are just two trivial adsorbing states for the
monomer-monomer and monomer-dimer models so the dimer-dimer model is
very different from these previously studied models. Rich kinetic
behaviors, resembling the ones of the deposition-evaporation
models\cite{stin,dhar}, can be envisioned.

\vskip 2cm

It is a pleasure to thank S.~Redner for fruitful discussions.
We gratefully acknowledge the financial support of the Swiss National
Foundation (to L.F.), and ARO grant DAAH04-93-G-0021 and NSF grant
DMR-9219845 (to P.L.K.).


\end{document}